\documentclass[prl,twocolumn,letterpaper,byrevtex,floatfix,showpacs,
superscriptaddress]{revtex4}

\usepackage{times,mathptmx}     
\usepackage{color}		
\usepackage[dvips]{graphicx}	
\usepackage{amsbsy}		
\usepackage{amsfonts}		
\usepackage{amsmath}		
\usepackage{amssymb}		
\usepackage{upgreek}		
\usepackage{color}
\usepackage{wasysym}
\usepackage{multirow}

\newcommand*{\uhecnu}{UHEC$\upnu$}

\newcommand*{\ie}{\emph{i.e.}}

\newcommand*{\sub}[1]{_{\mathrm{#1}}}
\renewcommand*{\sup}[1]{^{\mathrm{#1}}}

\newcommand*{\Es}{E_s}
\newcommand*{\Enu}{E_\nu}
\newcommand*{\Nnu}{N_\nu}
\newcommand*{\RL}{R\sub{\leftmoon}}

\newcommand*{\DeltaVC}{\Delta\sub{VC}}
\newcommand*{\thetaVC}{\theta\sub{VC}}
\newcommand*{\Ave}[1]{\mathinner{\left\langle{#1}\right\rangle}}

\begin{document}

\title{Prospects for Lunar Satellite Detection of Radio Pulses from\\
Ultrahigh Energy Neutrinos Interacting with the Moon}

\author{O.~St{\aa}l}
 \email{oscar.stal@tsl.uu.se}
 \affiliation{%
 Swedish Institute of Space Physics,
 P.\,O.\,Box 537, SE-751\,21 Uppsala,
 Sweden}%
 \affiliation{%
 High Energy Physics, Uppsala University,
 P.\,O.\,Box 535, SE-751\,21 Uppsala,
 Sweden}

\author{J.~E.~S.~Bergman}
 \affiliation{%
 Swedish Institute of Space Physics,
 P.\,O.\,Box 537, SE-751\,21 Uppsala,
 Sweden}%

\author{B.~Thid{\'e}}
 \affiliation{%
 Swedish Institute of Space Physics,
 P.\,O.\,Box 537, SE-751\,21 Uppsala,
 Sweden}%
\affiliation{LOIS Space Centre, V{\"a}xj{\"o} University,
 SE-351\,95 V{\"a}xj{\"o}, Sweden}

\author{L.~K.~S.~Daldorff}
 \affiliation{%
 Swedish Institute of Space Physics,
 P.\,O.\,Box 537, SE-751\,21 Uppsala,
 Sweden}%
 \affiliation{LOIS Space Centre, V{\"a}xj{\"o} University,
 SE-351\,95 V{\"a}xj{\"o}, Sweden}

\author{G.~Ingelman}
 \affiliation{%
 High Energy Physics, Uppsala University,
 P.\,O.\,Box 535, SE-751\,21 Uppsala,
 Sweden}%

\begin{abstract}
The Moon provides a huge effective detector volume for ultrahigh energy cosmic
neutrinos, which generate coherent radio pulses in the lunar surface layer due
to the Askaryan effect. In light of presently considered lunar missions, we
propose radio measurements from a Moon-orbiting satellite. First systematic
Monte Carlo simulations demonstrate the detectability of Askaryan pulses from
neutrinos with energies above $10^{20}$~eV, \emph{i.e.} near and
above the interesting GZK limit, at the very low fluxes predicted in different
scenarios.
\end{abstract}


\maketitle

Ultrahigh energy cosmic rays (UHECR) provide important information on
extreme astrophysical and cosmological processes.  Above the
Greisen-Zatsepin-Kuzmin (GZK) limit $\sim5\times10^{19}$~eV the universe
should be opaque over intergalactic scales to protons due to
$\mathrm{p}\upgamma\to \mathrm{N}\uppi$ where $\upgamma$ is a $2.7$~K
cosmic microwave background photon
\cite{Greisen:PRL:1966}.  The subsequent
decays $\uppi^+\to \upmu^+\upnu_{\mu}$ and $\upmu\to
\mathrm{e}\upnu_e\upnu_{\mu}$ give an isotropic flux of ``GZK
neutrinos'' \cite{Berezinsky&Zatsepin:PLB:1969,Engel&al:PRD:2001} which, if
detected, should resolve
\cite{Seckel&Stanev:PRL:2005} the apparent contradiction implied by
recent observations \cite{Takeda&al:APP:2003,Abbasi&al:PRL:2004} of
cosmic rays with energies above the GZK limit.

Weakly interacting ultrahigh energy neutrinos (\uhecnu) propagate unaffected
over cosmic distances, so their arrival directions point back to the
original sources.  Detection of neutrinos with energies well beyond the
GZK limit has also been suggested as a method to test cosmology through the
Z-burst process \cite{Weiler:PRL:1982}, in which the highest energy cosmic rays
would be produced following resonant interactions $\nu \bar{\nu}\to Z^0$
of {\uhecnu} with cosmological relic neutrinos.
In addition to these mechanisms, UHECR and {\uhecnu} could be produced in the
decay of super-heavy relic "X" particles originating \emph{e.g.}\ from
topological defects (TD) associated with Grand Unified Theories
\cite{Bhattacharjee&al:PRL:1992}. 
Recent neutrino flux limits, in particular that from ANITA-lite
\cite{Barwick&al:PRL:2006}, constrain these models and allow only a
percent-level contribution to the UHECR from Z-bursts.

No \uhecnu\ has been observed with existing neutrino telescopes.
The next generation km$^3$-sized optical detector IceCube
\cite{Ahrens&al:APP:2004} is optimized for $10^{12}$--$10^{17}$~eV neutrinos. At
even higher energies, it seems more promising
to explore the Askaryan effect \cite{Askaryan:JETP:1962} of
Vavilov-\v{C}erenkov (VC) radio pulse emission from a charged particle shower in
a dielectric medium. The shower, initiated by the interaction of a high energy
particle, produces via secondary scattering in the medium a net charge excess
which radiates coherently for wavelengths longer than the shower
dimension. Hence, the power radiated in radio scales quadratically with the
initial particle energy and not linearly as in the optical region.

The extremely low flux of \uhecnu\ necessitates a huge effective target.
This can be achieved by the detection of Askaryan pulses from neutrinos
interacting with the upper layer (regolith) of the Moon, which
has been considered for Earth-based radio measurements
\cite{Dagkesamanskii&Zheleznykh:JETP:1989,Hankins&al:MNRAS:1996,
Alvarez-Muniz&Zas:2001,Gorham&al:PRL:2004,Beresnyak&al:AR:2005,
Scholten&al:APP:2006}, and is currently proposed for a lunar satellite mission
\cite{Gusev&al:CR:2006}. Here, we report Monte Carlo (MC) simulation results on
Askaryan pulse detectability with instruments on a satellite orbiting the Moon.

The primary neutrino-nucleon cross sections for deep inelastic charged current
(CC) and neutral current (NC) interactions are well-known and can be
extrapolated to higher energies \cite{Gandhi:APP:1996}. Both CC and NC
interactions initiate hadronic showers carrying $\Es=y\Enu$ of the total
neutrino energy $E_\nu$, with $\Ave{y}=0.2$ at the very highest energies
\cite{Gandhi:APP:1996}. As discussed below, we only consider hadronic showers at
this stage. 

The Askaryan radio emissions from showers in different dielectric media
have been parametrized
\cite{Zas&al:PRD:1992,Alvarez-Muniz&Zas:PLB:1997,Alvarez-Muniz&al:PRD:2003}
and the results for Silica---the main constituent of the lunar
regolith---have been validated in accelerator experiments
\cite{Gorham&al:PRE:2000,Saltzberg&al:PRL:2001}.  In the form given by
\cite{Scholten&al:APP:2006}, the spectral flux density $F$ in Jansky
($10^{-26}\text{ W}\text{m}^{-2}\text{Hz}^{-1})$ of the VC
radiation in a frequency band $\Delta\nu$ around $\nu$, and at an angle
$\theta$ to the shower axis, from a charged particle shower of total
energy $\Es$ inside the lunar regolith can be expressed as
\begin{align}
\label{eq:F}
\begin{split}
&F(R,\theta,\nu,\Es) = 1.89\times10^{9}\,e^{-Z^2}
 \left(\frac{\sin\theta}{\sin\thetaVC}\right)^2
 \left(\frac{\Es}{E_0}\right)^2 \\
 \quad&\times\left(\frac{\RL}{R}\right)^2
 \left(\frac{\nu}{\nu_0[1+(\nu/\nu_0)^{1.44}]}\right)^2\frac{\Delta\nu}{100\,
\text{MHz}}.
\end{split}
\end{align}
Here, $E_0=10^{20}$~eV, $Z=(n\cos\theta-1)/(\DeltaVC\sqrt{n^2-1})$,
${\nu_0=2.5}$~GHz is the decoherence frequency where the wavelength becomes
comparable to the transverse dimension of the shower, $\RL=1.738\times10^6$~m
the lunar radius, $R$ the
distance from the source point in the regolith to the detector, and
$\DeltaVC=0.0302[\nu_0L(E_0)]/[\nu L(\Es)]$ radians
the angular spread around $\thetaVC$. The shower length
${L(E_s)~=12.7+2/3\log{(\Es/E_0)}}$ in units of radiation length
\cite{Scholten&al:APP:2006}. The detection threshold in eV is
\begin{align}
\label{eq:threshold}
\Es\sup{th} = 8.55\times10^{20}\frac{R}{\RL}\frac{\nu_0}{\nu}
 \left[1+\left(\frac{\nu}{\nu_0}\right)^{1.44}\right]
 \sqrt{\frac{N_\sigma^2 T\sub{noise}}{\Delta\nu A\sub{eff}}}
\end{align}
in terms of an effective antenna collection area $A\sub{eff}$,
a radio system noise temperature $T\sub{noise}$, and a minimum detectable
radio signal of $N_\sigma^2$ times the background noise.

Based on properties of lunar soil sample returns
\cite{Olhoeft&Strangway:EAPSL:1975}, we model the regolith down to
$100$~m depth as a homogeneous dielectric medium with a density of
$\rho=1.7\times10^3$~kg/m$^3$ and a radio attenuation length of
$l=\lambda/(2\pi n\tan\delta)$~m.  To allow for uncertainties in the
loss tangent, $\tan\delta$, due to metallic contaminants, we choose the
conservative value $\tan\delta=0.01$ which accounts for the available data.
Within uncertainties, this number is also consistent with predictions
made for the lunar bedrock layer \cite{Olhoeft&Strangway:EAPSL:1975}.
Since the attenuation in both materials can be consistently described by
a single value, the transition depth becomes unimportant, and our
simplified assumption of a deep, homogeneous regolith is justified for our
purposes.
\begin{figure}[t]
  \includegraphics[width=1.0\columnwidth,keepaspectratio]{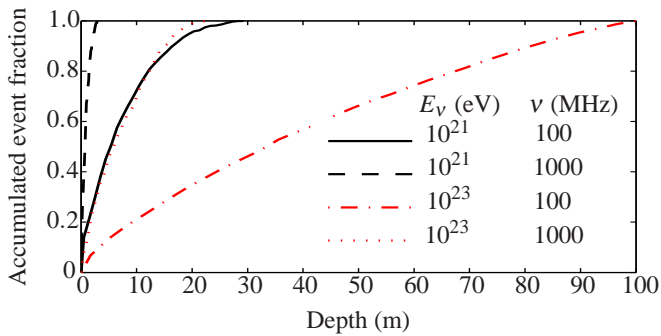}
\caption{(color online).  
Monte Carlo simulation results for the fraction of neutrino events (energy
$E_\nu$) above a given depth in the Moon, as detected by a tripole antenna
(frequency $\nu$) in a Moon-orbiting satellite at 100 km altitude.}
\label{fig:depth}
\end {figure}
To estimate the impact of choosing a different depth, below which no
neutrino events can be detected, we show in Fig.~\ref{fig:depth} the
accumulated fraction of detected events vs.\  depth, obtained from simulations
as detailed below. The effective depth
over which showers are detected is a function of both neutrino
energy and observation frequency. The full $100$~m depth only
contributes for the lowest frequencies and highest neutrino energies. Thus, the
curve for $\nu=100$~MHz, $E_\nu=10^{23}$~eV illustrates the maximum uncertainty.

For the calculation of the emission geometry we use $\thetaVC=55^\circ$,
corresponding to a constant regolith dielectric permittivity
$\varepsilon=3$ for the full radio frequency interval considered
\cite{Olhoeft&Strangway:EAPSL:1975}.  The total internal reflection
angle $\theta\sub{TIR}$ at the Moon-vacuum interface is complementary to
the VC angle.  Thus, for a narrow VC cone only emissions from neutrinos
which are upward-going with respect to the local surface will escape the
Moon.  For lower frequencies, when the cone is wider, total internal
reflection is not a significant problem since the interface transmission
rises rapidly to its maximum value just a few degrees off
$\theta\sub{TIR}$.  Rays covering the $\DeltaVC$ cone are propagated 
using geometrical optics, taking into account the effects of attenuation, 
refraction in the (locally flat) Moon-vacuum interface and internal reflection.

In order to estimate the optimum sensitivity in some generality, we use
two different approaches to define the radio sensor equipment.  In the
first case, an isotropic antenna with dipole characteristics is assumed,
representing a low-gain measurement of the complete electric field
vector.  This can be realized, for instance, using a tripole antenna
\cite{Compton:IEEEA:1981}.  The antenna length is
taken to be $\lambda/2$ at the highest frequency.  It will hence be
electrically short over the full bandwidth, which ensures nearly
single-mode operation.  For an isotropic measurement, the noise
temperature $T\sub{noise}$ is dominated at low frequencies by the
galactic background for which we use the simple model
$T\sub{gal}=1.5\times10^6\left(10\,\text{MHz}/\nu\right)^{2.2}$ K
\cite{Stremler:1991}.  

For higher frequencies, the predominant noise
contribution is from the radio receiver system and the satellite,
assumed at a nominal temperature $T_\mathrm{sys}=300$ K. In the second case, a perfectly beam-filling antenna array is assumed, \ie\ the
beam solid angle $\Omega$ equals the solid angle of the Moon. The corresponding
effective antenna area is then given by $A\sub{eff}\Omega=\lambda^2$ and the
physical size of the antenna array required to achieve this depends on both
frequency and altitude. The assumed directivity towards the lunar surface for
this case allows us to set $T\sub{noise}=T\sub{sys}=300$ K.

Unlike terrestrial measurements of RF transients, a lunar satellite
experiment suffers no atmospherics
\cite{Lehtinen&al:PRD:2004}. Likewise, anthropogenic noise, known to
interfere badly on Earth \cite{Hankins&al:MNRAS:1996}, is favorably
low in the lunar environment, in particular on the far side of the Moon
\cite{Alexander&al:AA:1975}.  For Gaussian thermal voltage fluctuations, the
single channel rate of spurious triggers,
exceeding a level $V_0$, is given by
$\Gamma_1=2\Delta\nu\times \mathrm{P}(|V|>V_0)=2\Delta\nu\times
\mathrm{erfc}(N_\sigma/ \sqrt{2})$. At $100$~MHz bandwidth, $N_\sigma^2=25$
gives $\Gamma_1=115$~s$^{-1}$. Requiring $n$-fold coincidence of independent
measurements in a time window $\tau$, this rate can be reduced to
$\Gamma_n=\tau^{n-1}\Gamma_1^n$.  In practice, multiple antennas, frequency
bands and/or polarizations are used to define the coincidence channels.  To
avoid technical details in this generic study, we simulate the detection system
using an effective description with full rejection of thermal events and the
sensitivity of a single channel. A threshold $N_\sigma^2=25$ is used,
corresponding to a realistic number for each channel in coincidence (cf.
\cite{Barwick&al:PRL:2006}).

For an isotropic neutrino flux $\Phi_\nu(E)$ the number of detected events in
the energy interval $[E_1,E_2]$, during time $\Delta t$, is 
\begin{align}
\Nnu = \Delta t\int_{E_1}^{E_2}\Phi_\nu(E)
 \alpha(E)\,\mathrm{d}E
\label{eq:N}
\end{align}
where the aperture function $\alpha(E)$ describes the total experimental
efficiency as a function of energy. Including the neutrino interaction, radio
wave propagation effects and the receiver noise, it represents an equivalent
detector area and effective solid angle for which 
$100\%
$ of the incident neutrino flux is detected. 
This complicated aperture is best studied by MC simulations.
For zero detected events (signal and background), a
model independent limit \cite{Lehtinen&al:PRD:2004} on a sufficiently smooth
neutrino flux function can be deduced by inverting (\ref{eq:N}) over an energy
interval $\mathrm{d}E\sim E$ to obtain 
\begin{align}
E \Phi_\nu(E)\le\frac{s\sub{up}}{\alpha(E)\Delta t}
\label{eq:lim}
\end{align}
with the Poisson factor $s\sub{up}=2.3$ for a limit with 
$90\%$ CL. 

\begin{figure}[t]
  \includegraphics[width=1.\columnwidth]{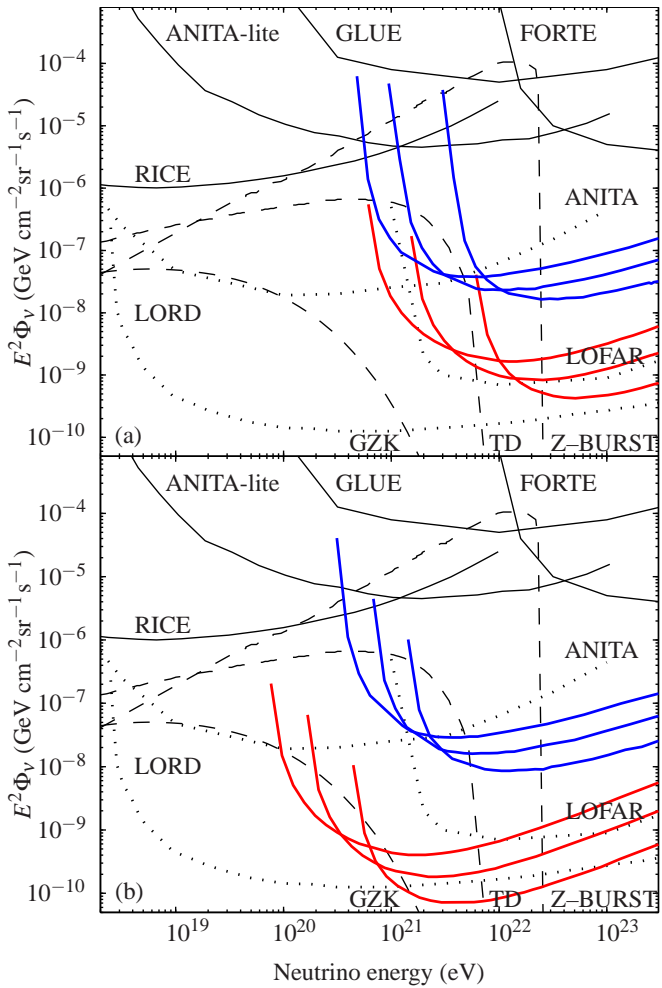}
  \caption{
  (color online).  $E^2$-weighted flux of \uhecnu.  Solid (color) curves
  show the projected detection limits from
  Eq.~(\ref{eq:lim}), based on one year of satellite measurements with
  (a) a single tripole antenna and (b) a beam-filling antenna 
  for frequencies of $100$ MHz (lower set of curves)
  and $1000$ MHz (upper set of curves).  Within each set, the curves
  from top to bottom are for satellite altitudes $H$ of $100$, $250$ and
  $1000$~km, respectively.  Dashed lines show predicted fluxes from the
  GZK process \cite{Engel&al:PRD:2001} (consistent with the
  Waxman-Bahcall bound $5\times 10^{-8}$
  \cite{Waxman&Bahcall:PRD:1999}), Z-bursts and Topological Defects
  (TD) \cite{Semikoz&Sigl:JCAP:2004}.  Thin solid lines show
  current flux limits from ANITA-lite \cite{Barwick&al:PRL:2006}, RICE
  \cite{Kravchenko&al:PRD:2006}, GLUE \cite{Gorham&al:PRL:2004} and
  FORTE \cite{Lehtinen&al:PRD:2004}. Dotted lines show predicted
  sensitivities for ANITA \cite{Barwick&al:PRL:2006}, LOFAR
  \cite{Scholten&al:APP:2006} and LORD \cite{Gusev&al:CR:2006}.
 }
 \label{fig:difflim}
\end {figure}
To map out the most favorable conditions for {\uhecnu} detection we
perform MC simulations of the aperture for several values of the center
frequency $\nu$ and 
altitude $H$ for a satellite in stationary, circular orbit. The frequency
range is limited to $100$--$1000$ MHz, since the narrower VC cone strongly
disfavors observations closer to the decoherence frequency $\nu_0$.
Around each center frequency, a symmetric bandwidth of $\pm 50$ MHz is
assumed.  The altitudes range from $100$ to $1000$ km. For the beam-filling
case, with $A_\mathrm{eff}$ depending on altitude, also the distance to Earth is
considered for comparison.  At energies
two orders of magnitude above threshold, where the detection starts to
become fully efficient, the apertures reach $10^3$--$10^6$ km$^2$sr for
moderate altitudes. 

Converting the resultant apertures to limits on the neutrino flux
according to Eq.~(\ref{eq:lim}), one limit curve is obtained for each
parameter pair $(\nu, H)$. Fig.~\ref{fig:difflim} shows the results for an
effective
observation time of one year, the minimum duration we
consider for a lunar satellite mission. The general trend is
that lower frequency observations yield more stringent flux limits as
a result of the increased angular spread 
$\DeltaVC$ of the Askaryan radiation and the longer radio attenuation
length in the regolith for lower frequencies. The optimum altitude is less
clear. From geometry, the accessible aperture increases with altitude as
the visible surface area of the Moon increases, giving stricter
limits.  However, the threshold energy also increases, which results in
the successive shift towards higher energies of the limit curves.
As shown in Fig.~\ref{fig:difflim}, our flux limits would improve
substantially over the existing ones.
They would also be competitive, and in some energy regions even better than,
the estimated limits that may be obtained by ANITA
\cite{Barwick&al:PRL:2006} and LOFAR \cite{Scholten&al:APP:2006}.
Predictions with both higher sensitivity and lower threshold can also be
obtained for a lunar satellite experiment, as indicated by the LORD
curve \cite{Gusev&al:CR:2006}, assuming that more elaborate antenna setups
can be realized.

Since the threshold energy and aperture cannot be varied independently,
optimum satellite parameters can only be judged with respect to a
specific neutrino flux model. Based on the fluxes from models for GZK,
Z-bursts and topological defects, Table \ref{tab:rates} presents integral event
rates calculated from Eq.~(\ref{eq:N}). For the Z-burst rates, we
conservatively rescale the original flux \cite{Semikoz&Sigl:JCAP:2004} by a
constant factor to conform with the integral bound $E^2\Phi_\nu \leq
1.6\times10^{-6}$ GeV cm$^{-2}$ sr$^{-1}$ s$^{-1}$ from ANITA-lite
\cite{Barwick&al:PRL:2006}. The rates thus obtained confirm that low frequency
observations are most efficient. It is evident that GZK neutrinos cannot be
detected in the single tripole case, whereas neutrinos from the other sources
can. Here, the lower threshold energy associated with lower altitudes is
strongly favorable for the detection of TD neutrinos, while detection of Z-burst
neutrinos gains more from the increased aperture at higher altitudes.

\begin{table}
\centering
 \caption{Event rates for satellite observations at
 frequencies $\nu$ and altitudes $H$ based on neutrino fluxes from the GZK
process \cite{Engel&al:PRD:2001}, a model for topological defects (TD, with
$M_X=2\times10^{13}$ GeV) and Z-bursts ($m_\nu=0.33$ eV, consistent with
cosmological data \cite{Semikoz&Sigl:JCAP:2004}, rescaled by a factor
$1.5\times10^{-2}$ to satisfy current flux limits).}         
\begin{tabular}{|cc|c|c|c|}
	\hline       
         \multicolumn{2}{|c|}{\bf Satellite parameters} &
\multicolumn{3}{c|}{\bf Integral event rates (yr$^{-1}$)} \\         
         $\nu$ (MHz) & $H$ (km) & GZK & TD & Z-burst \\ 
         \hline
         \multicolumn{2}{|l|}{\bf{(a) Tripole case}} &  & & \\
         100 & 100 & $4.9\times10^{-2}$ & $2.2\times10^2$ & $3.3\times10^3$ \\
         100 & 1000 & $<10^{-5}$ & 0.0 & $4.1\times10^3$ \\
         1000 & 100 & $6.8\times10^{-4}$ & 19 & 1.5$\times10^2$ \\
         1000 & 1000 & $<10^{-5}$ & $1.5\times10^{-1}$ & 1.9$\times10^2$ \\
         \multicolumn{2}{|l|}{\bf{(b) Beam-filling case}} &  & & \\
         100 & 100 & 21 & 6.5$\times10^3$ & 1.2$\times10^4$ \\
         100 & 1000 & 6.0 & 1.5$\times10^4$ & 7.6$\times10^4$ \\
         100 & Earth & 1.5 & 1.3$\times10^4$ & 1.8$\times10^5$ \\
         1000 & 100 & $2.3\times10^{-2}$ & 39 & 1.9$\times10^2$ \\
         1000 & 1000 & $1.2\times10^{-3}$ & 21 & 6.2$\times10^2$ \\
         1000 & Earth & $1.5\times10^{-4}$ & 7.4 & 9.9$\times10^2$ \\
         \hline        
\end{tabular}
\label{tab:rates}
\end{table}
 
In the beam-filling case the antenna gain pattern is adapted to the lunar disc,
resulting in higher event rates, but they are even more affected by the choice
of observation frequency.  
The suppressed galactic noise pushes the optimum frequency to its lowest
possible value, adding to the effect of an increasing $\DeltaVC$. A low
frequency beam-filling setup would be able to detect also the GZK neutrinos. 
For the TD and Z-burst models, the event rates depend weakly on the altitude.

These relatively low rates can be increased through several possible
improvements.  For CC $\upnu_e$ events, an additional shower results from the
emerging electron, which carries on average 
$80\%$ of the total neutrino energy. However, for electrons of high energy 
($\gtrsim 8\times10^{14}$ eV \cite{Alvarez-Muniz&al:PRD:2006}) the cross section
to interact with the material is reduced due to the LPM effect
\cite{Landau:JPUSSR:1944}. The result is an elongated shower, with a reduced
$\DeltaVC$ \cite{Alvarez-Muniz&Zas:PLB:1997}, and hence smaller detection
probability. Still, the electron-initiated showers, from the expected one third
of $\upnu_e$ in the primary neutrino flux, may be further investigated, as may
secondary showers from CC $\upnu_\tau$ interactions.

In a more detailed modeling of the lunar composition, also the positive
contributions of the bedrock layer should be considered.  The higher density
means more
neutrino interactions per unit volume and showers that develop faster, which
gives larger $\DeltaVC$ and increased detection probability.  
The effects of surface roughness scattering and possibly diffuse scattering
at the rock--regolith interface should then also be considered.

To increase the experimental sensitivity to GZK neutrinos, the threshold
energy [Eq.~(\ref{eq:threshold})] must be lowered by increasing the
effective antenna area, or by decreasing $N_\sigma$.  This means using more
antennas in coincidence and, for the beam-filling case, each
with a narrower beam.  The sensitivity is limited by what is technically
feasible to use  on-board a lunar satellite.  

In conclusion, we have used MC simulations to demonstrate the
prospects for a Moon-orbiting satellite to detect, via the Askaryan
effect, ultrahigh energy cosmic neutrinos interacting in the Moon.  This
method can provide competitive, and in some respects better,
conditions compared to Earth-based experiments.  For a given lunar
mission, the incremental cost for this experiment will not be
excessive. For the two specific model cases we consider, neutrino energies above
$10^{20}$~eV can be covered.  The flux of GZK neutrinos, which originate from
well known Standard Model processes, can be discovered with suitable satellite
and antenna parameters. The \uhecnu\ from proposed exotic sources can be
observed even at a much lower flux than predicted, resulting in strict limits to
be set or a revolutionary discovery. 

\begin{acknowledgments}
We are grateful to V.\ A.\ Tsarev and the LORD collaboration for very
interesting and useful discussions. Financial support from the Swedish
Governmental Agency for Innovation Systems, the Swedish National Space
Board and the Swedish Research Council is gratefully acknowledged. 
The computer cluster used was funded via an IBM Shared University
Research (SUR) grant.
\end{acknowledgments}

\bibliographystyle{apsrev}
\bibliography{lunarneutrinos}

\end{document}